\documentclass[prd,twocolumn,showpacs]{revtex4}
\usepackage{graphicx}
\usepackage{dcolumn}
\begin{document}
\title{Pure SU(3) lattice gauge theory\\using operators and states}
\author{J. B. Bronzan}
\email{bronzan@physics.rutgers.edu}
\affiliation{Department of Physics and Astronomy\\
              Rutgers University\\	      
136 Frelinghuysen Rd., Piscataway, NJ 08823}
\date{\today}
\begin{abstract}
We study pure SU(3) gauge theory on a large lattice, using Schr\"odinger's 
equation.  Our approximate solution uses a basis of roughly 1000 states.  
Gauge invariance is recovered when the color content of the ground state is 
extrapolated to zero.  We are able to identify the gauge invariant 
excitations that remain when the extrapolation is performed.  In the weak 
coupling limit, we obtain promising results when we compare the excitation 
energies (masses) to known exact results, which we derive.  We discuss the 
application of our nonperturbative method in the regime where glueballs 
are present.
\end{abstract}
\maketitle
\section{\label{sec:level1}Introduction.}

Pure SU(3) gauge theory on the lattice can be formulated in Hamiltonian 
form.  Matrix elements of the Hamiltonian using any finite orthonormal 
basis can be used to form a Hamiltonian matrix, and the eigenvalues of 
this matrix, ordered in increasing energy, are each an upper limit to the 
corresponding eigenvalue of the original Hamiltonian.\cite{A}  If the basis 
is provided with parameters, these can be used to lower the approximate 
energies closer to their true values.

Variational calculations run into problems unless the basis 
states are carefully chosen.  This is particularly true when many 
degrees of freedom (DF) are involved.  The most serious problem is that the 
eigenvalues of the matrix may be useless because they lie far above the 
eigenvalues of the Hamiltonian and do not accurately portray the 
physical implications of the Hamiltonian, quantitatively or even 
qualitatively.  Another problem associated 
with many DF is that the basis size can become unmanageable.  If we work 
on a $L\times L\times L$ spatial lattice, there are $24L^3$ degrees of 
freedom.  (There are 3 links emerging from each site on the lattice, 
and 8 colors per link.)  When we introduce two states for each DF, as 
we will, we have a basis of $2^{24L^3}$ states.  Further cuts must be 
made.  As we do this, we also want to build symmetries of the 
Hamiltonian into the basis so that quantum numbers can be 
assigned to the eigenvalues of the matrix.  The foundation of a 
variational calculation is therefore the crafting of a well-chosen basis.

The main purpose of this paper is to construct a basis for pure SU(3) lattice 
gauge theory and demonstrate that it is well-chosen.  
In this Introduction we outline the issues that must be faced in the 
construction process, and how we cope with them.  In following sections 
we compare results obtained with our basis to known results, thus testing 
the adequacy of the basis.  This comparison can be made only in a limiting 
case, but that case imposes significant tests.  The methods we use can be 
applied away from the limit to the ``physical'' case, where the lattice is 
inhabited by glueballs.  We outline how to do this at the end of the 
paper.

The approach we take in this paper is clearly an approximation.  On the 
other hand, the customary evaluation of the path integral by Monte Carlo 
sampling can, in principle, be arbitrarily accurate as more computing 
resources are devoted to the project.  But that project has been ongoing 
for 30 years, using the most advanced computing facilities in the world.  
The payoff to our approach is that the computer resources required are 
very modest.

\subsection{The Hamiltonian}

We begin with a description of the elements required to write the 
Hamiltonian for pure SU(3) local gauge theory on a three dimensional 
lattice.  We label links on our lattice by the pair $({\bf s},m)$, 
where $\bf s$ is the site from which a link departs, with 
$0\le s_x,s_y,s_z\le L\equiv 0$, and the lattice is an $L\times L\times 
L$ cube.  $m=x$, $y$ or $z$ is the direction 
of the link.   The variables assigned to a link degree of freedom 
specify a point on the SU(3) manifold, a compact manifold of eight 
dimensions labelled by $a=1,\dots,8$, the color index.  In the Hamiltonian 
we use the fundamental representation matrix $U_{{\bf s},m}$, which depends 
on the point on SU(3) for link $({\bf s},m)$.  We also use the operators 
${\cal J}_{L,a}({\bf s},m)$ with $a=1,\dots,8$.  The action of 
${\cal J}_{L,a}$ on a representation of SU(3) is to left-multiply it by 
the $a$th generator matrix for that representatrion.  Similarly, 
${\cal J}_{R,a}$ acts to right-multiply.  ${\cal J}^2=\sum {\cal J}^2_{L,a}
=\sum{\cal J}^2_{R,a}$ multiplies by the quadratic Casimir operator for 
SU(3).  The Hamiltonian is then:
\begin{eqnarray}
\label{eq:ham}
H_A&=&\sum_{\bf s}H({\bf s}),\nonumber\\
H({\bf s})&=&\frac{g^2}{2}\sum_{m}{\cal J}^2({\bf s,m})
-\frac{1}{g^2}
\end{eqnarray}
\begin{equation}
\times\sum_{m<n}\left[{\rm Tr}\left(U_{{\bf s},m}
               U_{{\bf s+\hat m},n}U^{\dagger}_{{\bf s+\hat n},m}
               U^{\dagger}_{{\bf s},n}\right)
               +{\rm H.c.}-6\right].\nonumber
\end{equation}
$H_A$ is the Kogut-Susskind Hamiltonian.\cite{D}  It commutes with the 
generators of local gauge transformations at site ${\bf s}$:
\begin{eqnarray}
\label{eq:gauge}
Q_a({\bf s})&=&-{\cal J}_{L,a}({\bf s},x)-{\cal J}_{L,a}({\bf s},y)
-{\cal J}_{L,a}({\bf s},z)\nonumber\\
& &+{\cal J}_{R,a}({\bf s-\hat x},x)+{\cal J}_{R,a}({\bf s-\hat y},y)
\nonumber\\
& &+{\cal J}_{R,a}({\bf s-\hat z},z).
\end{eqnarray}

The Hamiltonian formulation of lattice gauge theory has been used for a 
number of studies, mostly based on loop expansions that differ from the 
approach we take here.  For a recent review, see \cite{J}.

\subsection{States for Each Link.}

The first step in building a basis is to choose states on the SU(3) 
manifold for each link.  To make the choice, we need to know where 
on the manifold the eigenfunctions of $H_A$ have support (are large), and 
this depends on the coupling $g$.  We begin with small $g$.  It is convenient 
to use the parametrization for the fundamental representation matrix
\begin{equation}
U=\exp(ig\lambda_aR_a/2)
\end{equation}
where $\lambda_a$ is a Gell-Mann matrix.\cite{B}  We next develop 
ascending power series in $g$ for the elements in the Hamiltonian 
\begin{eqnarray}
\label{eq:weakops}
{\cal J}_{L,a}&=&\frac{1}{ig}\frac{\partial}{\partial R_a}+\frac{i}{2}f_{abc}
R_b\frac{\partial}{\partial R_c}+\cdots,\nonumber\\
{\cal J}_{R,a}&=&\frac{1}{ig}\frac{\partial}{\partial R_a}-\frac{i}{2}f_{abc}
R_b\frac{\partial}{\partial R_c}+\cdots,\\
U&=&1+\frac{ig}{2}(\lambda_aR_a)-\frac{g^2}{8}(\lambda_aR_a)^2\cdots .
\nonumber
\end{eqnarray}
To study support, we examine a ``lattice'' consisting of one link.  To lowest 
order in $g$, the Hamiltonian is 
\begin{equation}
H'=\frac{g^2}{2}{\cal J}^2-\frac{1}{g^2}
\left[2{\rm Tr}U-6\right]=-\frac{1}{2}\frac{\partial^2}{\partial 
{\bf R}^2}+\frac{{\bf R}^2}{2}.
\end {equation}
If we treat $\bf R$ as an eight dimensional Euclidean vector, the ground 
state of $H'$ is 
\begin{equation}
\label{eq:onelink}
\psi=\exp(-{\bf R}^2/2).  
\end{equation}
The Euclidean approximation is justified 
when $g$ is small because then the group parameters $\alpha_a=gR_a$ are 
small where the wave function is large, and the fact that SU(3) is not 
Euclidean can be ignored.  At weak coupling, 
therefore, wave functions have support only near the group origin.  
As $g$ grows, the region on the manifold where the wave function is large 
expands.  At large $g$ we know from strong coupling expansions that wave 
functions have support everywhere on the manifold\cite{D}.  Therefore, we 
seek Gaussian-like states on the manifold, centered on the origin, with 
width that can be tuned as a variational parameter.

Gaussian-like states on the group are generated by the kernel of the 
heat equation\cite{C}.
\begin{equation}
\frac{d\phi}{dt}={\cal J}^2\phi.
\end{equation}
The weak coupling, Euclidean approximation demonstates that the kernel 
is Gaussian:
\begin{equation}
\label{eq:kernel}
\frac{d\phi}{dt}=\frac{1}{g^2}\nabla^2\phi,\qquad \langle{\bf R_1}|
e^{t\nabla^2/g^2}|{\bf R_2}\rangle=\frac{e^{-g^2({\bf R_1-R_2})^2/4t}}
{(4\pi t/g^2)^4}.
\end{equation}
We will use the scaled width variable $\tau=t/g^2$.  Comparing 
Eqs. (\ref{eq:onelink}) and (\ref{eq:kernel}), for one link at 
weak coupling $\tau=1/2$.  This is the order of magnitude for $\tau$ 
that we will find for a large lattice at weak coupling.

The heat kernel on SU(3) is\cite{E}
\begin{eqnarray}
\psi_{\alpha_1}(\alpha)&=&\langle\{\alpha\}|e^{-t{\cal J}^2}|\{\alpha_1\}
\rangle\\
&=& \sum_{p,q}d(p,q)e^{-t\lambda(p,q)}\chi^{(p,q)}(\{\alpha\}
\{\alpha_1\}^{-1}),\nonumber
\end{eqnarray}
where $\{\alpha\},\ \{\alpha_1\}$ are group parameters, $\chi^{(p,q)}$ is 
the character, $d(p,q)$ is the dimension of representation $(p,q)$, and 
$\lambda(p,q)$ is the Casimir eigenvalue.\cite{E,F}  Note that when $t$ 
is small (narrow Gaussian), many representations contribute.  This mirrors 
the situation in Euclidean space, where many harmonics are required to 
build up a narrow Gaussian.

In Eucliean space, Gaussians can be extended to a complete set of basis 
states by multiplication by polynomials or by differentiation, and the same 
can be done on SU(3).  The simplest of these states are
\begin{eqnarray}
\phi=\psi_{\alpha1}(\alpha)|_{\alpha_1=0},\\
\phi_a={\cal J}_{La}(\alpha_1)\psi_{\alpha_1}(\alpha)|_{\alpha_1=0}.
\nonumber
\end{eqnarray}
The orthogonality properties of these states are\cite{E}
\begin{eqnarray}
\langle\phi|\phi\rangle&=&S_1;\quad\langle\phi_a|\phi\rangle=0;\quad
\langle\phi_a|\phi_b\rangle=\delta_{ab}S_2;\nonumber\\
S_1&=&\sum_{p,q}d^2(p,q)e^{-2t\lambda(p,q)};\\
S_2&=&\frac{1}{8}\sum_{p,q}d^2(p,q)\lambda(p,q)e^{-2t\lambda(p,q)}.
\nonumber
\end{eqnarray}
Matrix elements of the operators in Eq. (\ref{eq:ham}) among states 
$\phi$ and $\phi_a$ are computed in Ref. \cite{E}.  The dynamics of 
SU(3) is invoked through the use of these matrix elements.

On each link, we allow single excitations to occur in each of the eight 
colors.  We exclude states that are {\it multiply} excited in the same 
color direction. Since each of the eight color directions on a link can 
be unexcited ($\phi$-like) or singly excited ($\phi_a$-like), our basis 
has $2^8=256$ states per link.  An important topic of the paper will be 
to test the adequacy of limiting the basis to 256 states on each link in 
this way.

States in which two color directions on a link are excited, designated 
$\phi_{ab}$, are studied in Ref. \cite{E}, but states in which 
$3,\dots, 8$ directions are excited have not been studied.  Extending the 
number of color directions excited on a link in the explicit manner of 
Ref. \cite{E} is laborious and will not be attempted in this paper.  The 
process is also ambiguous because the operator ${\cal J}_{La}$ used in 
the definition of $\phi_a$ and $\phi_{ab}$ can be replaced by others, like 
${\cal J}_{Ra}$.  Instead, in this paper we introduce states 
with ``multi-color'' excitations on a link, and matrix elements involving 
these states, in a straighforward way we describe in the next subsection.

\subsection{States on the Lattice}

The lattice basis constructed by putting each link in one of the 256 states 
we have introduced has $(256)^{3L^3}=2^{24L^3}$ elements, which is far 
too many to handle.  In addition, local degrees of freedom constitute an 
awkward basis for describing phenomena that range over many links.  
Continuum field theory instructs us how to depart from 
local degrees of freedom by recasting the theory in momentum space.  We 
adopt that method in this paper.  

We introduce operators $A_a({\bf s},m)$.  These are bosonic, satisfying 
the usual bosonic commutation rules.  The ground state 
$|0\rangle$ corresponds to the normalized state $|\phi\rangle/\sqrt{S_1}$ 
on a link, and $-iA_a^{\dagger}|0\rangle$ corresponds to the state 
$|\phi_a\rangle/\sqrt{S_2}$ on the link.  At this point we can describe 
how we represent a normalized state in which two distinct colors are 
excited on the same link; it is 
$(-iA_a^{\dagger})(-iA_b^{\dagger}|0\rangle$, where $a\ne b$.

We now follow field theory and introduce nonlocal states by passing to 
momentum space, introducing the bosonic operators 
\begin{equation}
\label{eq:momentum}
B_a({\bf k},m)=\frac{1}{L^{3/2}}\sum_{\bf s}e^{-2\pi i{\bf k\cdot} 
({\bf s+\hat m}/2)/L}A_a({\bf s},m).
\end{equation}

\subsection{The Basis Generator G}

We return to the requirement that the state on each link be required to 
be in one of 256 states.  Applications of operators $(-iA_a^\dagger)^k$ 
with $k\ge 2$ produce states beyond the ones we've introduced.  We refer 
to such states as ``overexcited''.  Overexcited states 
will appear unless we suppress them.  First we introduce an operator that 
measures the excitation of the color/link degrees of freedom.  It is
\begin{equation}
N=\sum_{{\bf s},m,a}A_a^{\dagger}({\bf s},m)A_a({\bf s},m),
\end{equation}
A color/link degree of freedom in the $k$-th bosonic state is an 
eigenfunction of $N$ with eigenvalue $k$.  Our method of minimizing the 
contribution of overexcited states to our basis is to control the 
expectation value of $N$ in our basis states.

With the introduction of $N$, we can begin to construct the basis.  First 
we write $H_A$ in terms of our bosonic operators.  Coefficients are chosen 
so that matrix elements on the basis of the two states for each color/link 
agree with matrix elements derived in Ref. \cite{E}.  Next we decompose 
$H_A$ into a term $H_{A,0}$, quadratic in $A$ and $A{^\dagger}$ and a 
remainder:
\begin{equation}
H_A=H_{A,0}+H_{A,1}
\end{equation}
Equipped with this decomposition, we construct a generator operator $G$, 
whose low-lying eigenstates will compose our basis.  
The first expression for $G$ is
\begin{equation}
G_A=H_{A,0}+\xi N.
\end{equation}
By limiting $G_A$ to quadratic terms, it becomes an operator that can 
be diagonalized.  Its eigenstates can be constructed, leading to an explicit 
basis.  We take $\xi$ to be a positive parameter, so the term $\xi N$ in 
$G_A$ reduces the contribution of overexcited states to the low energy 
spectrum of $G_A$, where the states in our basis are found. By 
changing $\xi$, we can adjust $\langle N\rangle$ for the eigenstates of 
$G_A$.  Define $N_{ave}$ to be the expectation of $N$ per color/link DF 
in the ground state of $G_A$:  $N_{ave}=\langle 0|N|0\rangle/(24L^3)$.  
Most of our simulations will be carried out with $N_{ave}=1/2$.  This small 
value suggests that the probability of finding a color/link to be overexcited 
is small, and it is.  We will present a formula giving the probability 
that a color/link is in {\it any} overexcited state, and the 
typical result (when $N_{ave}=1/2$) is about 1/8.  Therefore, overexcited 
states contribute to results computed in our basis, but not much.  We have 
some ability to check on the influence of overexcited states by changing 
$\xi$ and hence $N_{ave}$.  We do not want to make $N_{ave}$ too small, 
however, because we do not want to suppress the 256 states we 
allow for each link.  Another method of excluding overexcited states would 
be to map pure SU(3) onto a spin model, with each color/link being in either 
a ``spin up'' or ``spin down'' state.  This would eliminate overexcited 
states altogether, but it would deny us the methods of field theory, 
particularly the simple introduction of nonlocal degrees of freedom by 
transforming to momentum space.

To summarize our procedure, we have adopted a bosonic formalism.  This allows 
us to pass to momentum space and construct a basis that is nonlocal.  The 
price we pay is that there are contributions from overexcited bosonic states 
that are unwanted.  By taking our basis states to be eigenstates of the 
generator $G_A$, we can use the parameter $\xi$ to minimize and assess the 
effects of the overexcited states.

\subsection{Gauge Invariance}

We seek states having $Q_a({\bf s})|\psi\rangle=0$.  
There are three options for excluding non-gauge invariant states.  One is to 
rewrite the theory in terms of gauge invariant DF only.  Unfortunately, the 
resulting Hamiltonians are nonlocal and are not manifestly under 
rotations of the cubic lattice\cite{G,H}.  They are unsuited for 
approximation schemes.  The second option is to use Hamiltonian 
(\ref{eq:ham}) but allow only gauge invariant states in the basis.  This 
is done when loop expansions are employed.\cite{J}  However, sound arguments 
have led us adopt the eigenstates of $G_A$ as basis 
states, and these states are not gauge invariant.  

The third option is to compute with a nonzero expectation value, per site, 
of the group quadratic Casimir operator: 
$Q_{ave}^2=\langle 0|Q^2|0\rangle/L^3$.  Here $|0\rangle$ is the ground 
state of $G_A$, and  
\begin{equation}
Q^2=\sum_{{\bf s},a}Q_a({\bf s})^2.
\end{equation}
This is the option we adopt.  We will calculate results for several values 
of $Q_{ave}^2$ and extrapolate to the limit $Q_{ave}^2=0$ at 
the end.  This was done successfully for the ground state in Ref. \cite{A}, 
and we use the same method here.  Note that $Q_a({\bf s})^2$ is a positive 
operator, so when $Q_{ave}^2=0$, gauge invariance is imposed at every site 
on the lattice. 

Once we include gauge non invariant states in our basis, we want to control 
their number.  This can be accomplished by adding a term to $G$ that 
raises the eigenvalue of such states.  To do this, we decompose $Q^2$ 
into quadratic and higher terms, just as we 
did for $H_A$
\begin{equation}
Q^2=Q_0^2+Q_1^2.
\end{equation}
We add term proportional to $Q_0^2$ to our generator, which we call 
$G_B$ after the modification.
\begin{equation}
\label{eq:generator}
G_B=G_A+\eta g^2Q_0^2=H_{A,0}+\xi N +\eta g^2Q_0^2.
\end{equation}
The factor $g^2$ is included because $g^2Q^2_{ave}$ is finite as 
$g\rightarrow 0$.

Finally we introduce $H_B$, the Hamiltonian we use in the rest of this paper.
\begin{equation}
\label{eq:finalham}
H_B=H_A+\eta g^2Q^2.
\end{equation}
$H_A$ and $H_B$ have the same set of physical, gauge invariant 
eigenstates.  However, the non-gauge invariant eigenstates of $H_B$ are 
raised in energy by the positive operator $\eta g^2Q^2$.  By varying $\eta$ 
we can control the number of gauge non invariant states in the low energy 
spectrum.

Our basis states are the low-lying eigenstates of generator $G_B$.  To 
diagonalize $G_B$ we must remove operators of the form 
$B^\dagger B^\dagger$ and $BB$.  This is accomplished by means of the 
Bogoliubov transformation to new bosonic operators $E$:
\begin{eqnarray}
\label{eq:bogoliubov}
\lefteqn{B_a({\bf k},m)=}\\
& &\cosh\theta({\bf k})E_a({\bf k},m)-\sinh\theta
({\bf k})E^{\dagger}_a({\bf L-k},m),\nonumber
\end{eqnarray}
with $\theta({\bf k})=\theta({\bf L-k})$ and ${\bf L}=(L,L,L)$.  
(Directional indices for $\theta$ will be introduced later, after we 
transform to the polarization basis.)  We use the vacuum 
state of $G_B$ after the Bogoliubov transformation to determine the 
expectation values $N_{ave}$ and $Q_{ave}^2$.  These parameters are set 
primarily by choice of $\xi$ and $\tau$.  $\eta$ is used to set the number 
of non gauge invariant states in the low energy spectrum of $G_B$.

\subsection{Group Theory, Degeneracy and Spin}

Hamiltonian (\ref{eq:ham}) is invariant under the group of rotations of 
the cubic lattice.  Associated with this symmetry is degeneracy; states 
belonging to different rows of the same irreducible representation of the 
group are degenerate.  Such degeneracy complicates computer work.  It can 
be reduced by forming linear combinations of our basis states that 
belong to the same row of an irreducible representation of the group:
\begin{equation}
\label{eq:group}
|\psi\rangle=\sum_{R}{\cal D}^{(J)*}_{\mu\nu}(R)U(R)|a\rangle,
\end{equation}
where ${\cal D}^{(J)}$ is an irreducible representation of the group, 
and $U(R)|a\rangle$ is the image of $|a\rangle$ under a rotation of the 
cube.

The group of rotations of the cube has 24 elements\cite{I}.  It has five 
irreducible representations (because there are five classes), which we 
label by their dimensions: $1_A$, $1_B$, $2$, $3_A$ and $3_B$.  (The sum 
of the squares of these dimensions is the order of the group, implying 
that Eq. (\ref{eq:group}) provides 24 states for each $|a\rangle$.)

An additional reason to take account of the symmetry group is that we wish 
to identify the spins of the states we find.  Since our lattice 
breaks the O(3) symmetry of continuum field theory, this identification 
is not obvious.  Note, however that each of the 24 rotations 
of the cube corresponds to an element of O(3).  The identity element 
corresponds to a rotation by $0^\circ$, and each of the other elements 
corresponds to a rotation by $90^\circ$, $120^\circ$ or a multiple of these 
about an axis through the center of the cube.  It follows that every 
irreducible representation of O(3) generates a representation of the group 
of the cube, which is generally reducible.  The first few Clebsch-Gordon 
series giving the content of O(3)-induced representations are
\begin{eqnarray}
({\rm spin 0})&\longrightarrow&1_A,\nonumber\\
({\rm spin 1})&\longrightarrow&3_A,\\
({\rm spin 2})&\longrightarrow&2\oplus 3_B,\nonumber\\
({\rm spin 3})&\longrightarrow&1_B\oplus 3_A\oplus 3_B.\nonumber
\end{eqnarray}
In these four cases, a representation of the group of the cube is 
identified with a spin.  However, an eigenstate belonging to the 
$2$ representation of the cube can be associated with spin 2 only if there 
is an eigenstate belonging to the $3_B$ of nearly the 
same energy.  If this is not the case, it may not be possible to identify 
the $2$ eigenstate with a continuum state.

In this paper we will study spin 0 states.  ${\cal D}(R)=1$ for this 
case, and
\begin{equation}
|\psi\rangle=A\sum_R U(R)|a\rangle
\end{equation}

\section{\label{sec:level2}Weak Coupling}

We need a known reference theory which we can use to evaluate the adequacy 
of the approximations laid out in Sec.~\ref{sec:level1}.  For this pupose 
we use the weak coupling limit of the theory.  This is the limit in which 
the correlation length has grown beyond the lattice length 
$L$.  This regime is reached for couplings $g$ {\it below} those in the 
scaling region in which glueballs are dynamically contained within the 
lattice.  The weak coupling limit is unphysical, but it provides a 
nontrivail test of the numerical accuracy of calculations using the basis 
produced by state generator $G_B$.  
In the weak coupling limit we can also test how well we can spot 
gauge-invariant excitations in a congeries of states that are 
excitations of $Q^2$.  Finally, we can test the extrapolation to 
$Q_{ave}^2=0$. 

To obtain the weak coupling form of the theory, use the weak coupling 
operators of Eq. (\ref{eq:weakops}) in the Hamiltonian, Eq. (\ref{eq:ham}).  
We retain only the leading terms in powers of $g$, which produces a 
free field theory:
\begin{eqnarray}
\label{eq:osc}
H_W&=&-\frac{1}{2}\sum_{{\bf s},m,a}\frac{\partial^2}{\partial 
R_a({\bf s},m)^2}\\
& &+\frac{1}{2}\sum_{{\bf s},m<n,a}[R_a({\bf s},m)+
R_a({\bf s+\hat m},n)\nonumber\\
& &-R_a({\bf s+\hat n},m)-R_a({\bf s},n)]^2.\nonumber
\end{eqnarray}
The generators of local gauge transformations are
\begin{eqnarray}
\label{eq:weakgauge}
Q_{W,a}({\bf s})&=&\frac{1}{ig}\left[-\frac{\partial}{\partial R_a({\bf s},x)}
-\frac{\partial}{\partial R_a({\bf s},y)}\right.\nonumber\\
& &-\frac{\partial}{\partial R_a({\bf s},z)}+\frac{\partial}{\partial 
R_a({\bf s-\hat x},x)}\nonumber\\
& &+\left. \frac{\partial}{\partial R_a({\bf s-\hat y},y)}+\frac{\partial}
{\partial R_a({\bf s-\hat z},z)}\right].
\end{eqnarray}

We diagonalize this theory by introducing bosonic operators.
\begin{eqnarray}
\label{eq:weakops2}
R_a({\bf s},m)&=&\sqrt{\sigma}[A_a({\bf s},m)+A^\dagger_a({\bf s},m)],\\
\frac{\partial}{\partial R_a({\bf s},m)}&=&
\frac{1}{2\sqrt{\sigma}}[A_a({\bf s},m)-A^\dagger_a({\bf s},m)].\nonumber
\end{eqnarray}
$\sigma$ is an arbitrary parameter, and physical consequences of 
Hamiltonian (\ref{eq:osc}) do not depend on $\sigma$. Next we introduce 
bosonic operators in momentum space:         
\begin{equation}
A_a({\bf s},m)=\frac{1}{L^{3/2}}\sum_{{\bf k}}e^{2\pi i{\bf k\cdot 
(s+\hat m/2)}/L}B_a({\bf k},m)
\end{equation}
We find that 
\begin{eqnarray}
\lefteqn{R_a({\bf s},m)+R_a({\bf s+\hat m},n)-R_a({\bf s+\hat n},m)
-R_a({\bf s},n)}\nonumber\\
&=&2i\sqrt{\sigma}\sum_{\bf k}e^{2\pi i{\bf k\cdot s}/L}
e^{i\pi(k_m+k_n)/L}\\
& &\times\left[\sin\left(\frac{\pi k_m}{L}\right)W_a({\bf k},n)
-\sin\left(\frac{\pi k_n}{L}\right)W_a({\bf k},m)\right],\nonumber
\end{eqnarray}
where $W_a({\bf k},m)=B_a({\bf k},m)-B^\dagger_a({\bf L-k},m)$.
This may be written concisely in terms of the unit longitudinal polarization 
vector
\begin{eqnarray}
\lefteqn{{\bf\hat\epsilon}^{(3)}({\bf k})=\omega({\bf k})\left[{\bf\hat x}
\sin\left(\frac{\pi k_x}{L}\right)+{\bf\hat y}\sin\left(\frac{\pi k_y}{L}
\right)\right. }\\
& &+\left. {\bf\hat z}\sin\left(\frac{\pi k_z}{L}\right)\right],\qquad
\omega^2({\bf k})=\sum_i\sin^2\left(\frac{\pi k_i}{L}\right),\nonumber
\end{eqnarray}
and the vector
\begin{equation}
{\bf W}_a({\bf k})={\bf\hat x}W_a({\bf k},x)+{\bf\hat y}
W_a({\bf k},y)+{\bf\hat z}W_a({\bf k},z).
\end{equation}
as
\begin{eqnarray}
& &R_a({\bf s},m)+R_a({\bf s+\hat m},n)-R_a({\bf s+\hat n},m)
-R_a({\bf s},n)\nonumber\\
& &=\pm\frac{2i\sqrt{\sigma}}{L^{3/2}}\sum_{{\bf k}}e^{2\pi i{\bf k\cdot s}/L}
e^{i\pi(k_m+k_n)/L}\nonumber\\
& &\times\ \omega({\bf k})\left[{\bf\hat\epsilon}^{(3)}
{\bf\times W}_a({\bf k})\right]_p,
\end{eqnarray}
with $m$, $n$, $p$ a permutation of x, y, z.  The sign of the term is 
positive (negative) for an even (odd) permutation. 
Introduce the real transverse polarization vectors 
${\bf\hat\epsilon}^{(1)}({\bf k})$ and 
${\bf\hat\epsilon}^{(2)}({\bf k})$ so that the three 
polarization vectors form an orthonormal right-handed basis.  The vectors 
are chosen so that ${\bf\hat\epsilon}^{(s)}({\bf k})
={\bf\hat\epsilon}^{(s)}({\bf L-k})$.  The polarization basis degrees of 
freedom are 
\begin{equation}
B^{(s)}_a({\bf k})=\sum_{m}{\bf\hat\epsilon}^{(s)}_{m}({\bf k)}
B_a({\bf k},m).
\end{equation}
Then
\begin{eqnarray}
\frac{1}{2}\sum_{{\bf s},m<n,a}[R_a({\bf s},m)+R_a({\bf s+\hat m},n)
\nonumber\\
-R_({\bf s+\hat n},m)-R_a({\bf s},n)]^2\nonumber\\
=2\sigma\sum_{{\bf k},a}\sum_{s=1}^2 \omega^2({\bf k})
W^{(s)}_a({\bf k})W^{(s)\dagger}_a({\bf k}).
\end{eqnarray}
Altogether, the weak coupling Hamiltonian is
\begin{eqnarray}
\label{eq:wkham}
H_W=\frac{1}{8\sigma}\sum_{{\bf k},a}\sum_{s=1}^3[B^{(s)}_a({\bf k})
+B_a^{(s)\dagger}({\bf L-k})]\nonumber\\
\times[B^{(s)\dagger}_a({\bf k})+B_a^{(s)}({\bf L-k})]\\
+2\sigma\sum_{{\bf k},a}\sum_{s=1}^2\omega^2({\bf k})[B^{(s)}_a({\bf k})
-B_a^{(s)\dagger}({\bf L-k})]\nonumber\\
\times[B^{(s)\dagger}_a({\bf k})-B_a^{(s)}({\bf L-k})]\nonumber
\end{eqnarray}

We now use the Bogoliubov transformation, Eq. (\ref{eq:bogoliubov}), on the 
transverse modes, $s=1,2$. The parameter is $2\theta^{(s)}({\bf k})
=-\ln [4\sigma \omega({\bf k})]$.  The ``transverse'' mode Hamiltonian 
becomes
\begin{equation}
H_{W,T}=L^3{\cal E}_0+2\sum_{{\bf k},a}\sum_{i=1}^2
\omega({\bf k})E^{(s)\dagger}_a({\bf k})E^{(s)}_a({\bf k}),
\end{equation}
and the ground state energy density is
\begin{equation}
{\cal E}_0=\frac{16}{L^3}\sum_{\bf k}\omega({\bf k}).
\end{equation}
${\cal E}_0$ depends weakly on the lattice size.  When $L=11$, 
${\cal E}_0=19.0999$, while when $L=31$, ${\cal E}_0=19.1008$.

The longitudinal modes ($s=3$) cannot be brought into this form.  A glance at 
Eq. (\ref{eq:osc}) shows that the free field theory is a set of coupled 
oscillators.  However the ``restoring'' term vanishes for the longitudinal 
modes, which therefore have a continuous spectrum of eigenstates 
corresponding to free motion.  The longitudinal mode Hamiltonian is
\begin{eqnarray}
H_{W,L}=\frac{1}{8\sigma}\sum_{{\bf k},a}\left[B^{(3)}_a({\bf k})+
B^{(3)\dagger}_a({\bf L-k})\right]\nonumber\\
\times\left[B^{(3)\dagger}_a({\bf k})+B^{(3)}_a({\bf L-k})\right].
\end{eqnarray}
Define four hermitian operators associated with mode $\bf k$ in this sum:
\begin{eqnarray}
\rho_{1,a}&=&\frac{1}{2i}[-B^{(3)}_a({\bf k})+
B^{(3)\dagger}_a({\bf k})\\
&&-B^{(3)}_a({\bf L-k})+B^{(3)\dagger}_a({\bf L-k})],\nonumber\\
\rho_{2,a}&=&\frac{1}{2}[B^{(3)}_a({\bf k})
+B^{(3)\dagger}_a({\bf k})\nonumber\\
&&-B^{(3)}_a({\bf L-k})-B^{(3)\dagger}_a({\bf L-k})],\nonumber\\
\pi_{1,a}&=&\frac{1}{2}[B^{(3)}_a({\bf k})
+B^{(3)\dagger}_a({\bf k})\nonumber\\
&&+B^{(3)}_a({\bf L-k})+B^{(3)\dagger}_a({\bf L-k})],\nonumber\\
\pi_{2,a}&=&\frac{1}{2i}[B^{(3)}_a({\bf k})
-B^{(3)\dagger}_a({\bf k})\nonumber\\
&&-B^{(3)}_a({\bf L-k})+B^{(3)\dagger}_a({\bf L-k})].\nonumber
\end{eqnarray}
These operators satisfy canonical commutation rules.
\begin{eqnarray}
\left[\rho_{1,a},\rho_{2,b}\right]=0,
\qquad\left[\pi_{1,a},\pi_{2,b}\right]=0,\nonumber\\
\left[\rho_{m,a},\pi_{n,b}\right]=i\delta_{m,n}\delta_{a,b}.
\end{eqnarray}
In terms of these operators, the longitudinal Hamiltonian is
\begin{eqnarray}
\lefteqn{H_{W,L}=\frac{1}{8\sigma}\sum_{{\bf k},a}
[B^{(3)}_a({\bf k})+B_a^{(3)\dagger}({\bf L-k})]}\nonumber\\
& &\times[B^{(3)\dagger}_a({\bf k})
+B_a^{(3)}({\bf L-k})]\\
& &=\frac{1}{8\sigma}\sum_{{\bf k},a}\{[\pi_{1,a}({\bf k})]^2
+[\pi_{2,a}({\bf k})]^2\}.\nonumber
\end{eqnarray}

Thus we need eigenstates of $\pi_{1,a}$ and $\pi_{2,a}$.  Since 
$[\pi_{1,a},\pi_{2,a}]=0$, we can look for common eigenfunctions:
\begin{equation}
\pi_{1,a}|p_1,p_2\rangle=p_1|p_1,p_2\rangle,\qquad
\pi_{2,a}|p_1,p_2\rangle=p_2|p_1,p_2\rangle.
\end{equation}
In Fock space the eigenfunction is
\begin{equation}
|p_1,p_2\rangle=\sum_{m,n=0}^{\infty}\frac{c_{m,n}}{m!n!}
[B^{(3)\dagger}_a({\bf k})]^m
[B^{(3)\dagger}_a({\bf L-k})]^n|0\rangle,
\end{equation}
with
\begin{eqnarray}
c_{m,n}&=&\frac{(i)^{m+n}}{\sqrt{\pi}}\int_{-\infty}^{\infty}d\lambda
e^{-\sqrt{2}\lambda p_1}\phi_m(\lambda+p_2/\sqrt{2})\\
& &\times\phi_n(\lambda-p_2/\sqrt{2}).\nonumber
\end{eqnarray}
The functions in this expression belong to a set of orthonormal harmonic 
oscillator eigenfunctions:
\begin{equation}
\phi_m(\lambda )=\frac{1}{\sqrt{m!\sqrt{\pi}}}\left[
\frac{1}{\sqrt{2}}\left(\lambda -\frac{d}{d\lambda}\right)\right]^m
e^{-\lambda^2/2}
\end{equation}
The eigenvalues are continuous, with $-\infty<p_1,p_2<\infty$, and the 
states are orthonormal:
\begin{equation}
\langle p_1,p_2|\tilde p_1,\tilde p_2\rangle=\delta(p_1-\tilde p_2)
\delta(p_2-\tilde p_2). 
\end{equation}

The mean excitation is $N_{ave}=\infty$ for the weak coupling theory.  
(A simple way to see this is provided by Eq. (\ref{eq:ave}).  First set 
$\eta=0$ and then examine the limit as $\xi\rightarrow 0$.)  This means 
that overexcited states contribute significantly to the eigenstates.  
This result highlights an important difference between the ground state of 
the weak coupling theory and the ground state of generator $G_B$, where we 
set $N_{ave}=0.5$ in most of our simulations.  Our approximation scheme 
is viable only if we can demonstrate the accuracy of calculations using 
our basis of states, despite this difference in $N_{ave}$.

The operator $Q^2$ may also be written in terms of the longitudinal 
operators we have introduced.
\begin{eqnarray}
\label{eq:queue}
\lefteqn{Q^2=\frac{1}{g^2\sigma}\sum_{{\bf k},a}\omega^2({\bf k})
[B^{(3)}_a({\bf k})+B_a^{(3)\dagger}({\bf L-k})]}\nonumber\\
& &\times[B^{(3)\dagger}_a({\bf k})
+B_a^{(3)}({\bf L-k})]\\
& &=\frac{1}{g^2\sigma}\sum_{{\bf k},a}\omega^2({\bf k})
\{[\pi_{1,a}({\bf k})]^2+[\pi_{2,a}({\bf k})]^2\}.\nonumber
\end{eqnarray}

Since $\omega^2\le 3$, we see that the ground state energy density 
for general $Q_{ave}^2$ is
\begin{equation}
\label{eq:gsenergy}
{\cal E}(Q_{ave}^2)={\cal E}_0+\frac{g^2Q_{ave}^2}{24}
\end{equation}
Energy levels above the ground state are comprised of a continuous 
spectrum of excitations of $Q^2$ and a discrete 
spectrum of gauge-invariant excitations whose energies are sums of 
$2\omega ({\bf k})$ for different values of ${\bf k}$.

Pure SU(3) is believed to exhibit color confinement, but our free field 
theory does not, nor will the theories we propose for $g^2>0$.  We contend 
that this is acceptable as long as we compute low energy properties of the 
theory such as glueball masses.  (Computations of the spectrum of charmonium 
show that the confinement potential plays a minor role there.) However, we 
must take account of color confinement by always contracting color indices 
$a$ into guage-invariant combinations.  For glueballs at rest, we will use 
a basis of gluon pair states of the form 
$\sum_aE_a^{(s)\dagger}({\bf k})E_a^{(s)\dagger}({\bf L-k})|0\rangle$.  
(In this paper we replace negative wave vectors $\bf -k$ by ${\bf L-k}$ 
so that all vector components lie in the interval $0\le k_\mu < L$.)
It will be seen that such states arise naturally in the course of 
calculation.  Thus, low-lying gauge invariant states at rest have energies 
(masses) $4\omega({\bf k})$.  This is the result we hope to replicate 
using our approximate eigenstates of $H_B$.

The spectrum of transverse excitations is highly degenerate.  Some of these 
degeneracies depend on familiar trigonometric identities and the 
trigonometric form of $\omega({\bf k})$.  For example, when $L$ is even, 
and $k_y=L/2\pm k_x$, $\omega^2=1+\sin^2(\pi k_z/L)$ for {\it any} $k_x$.  
When $L$ is divisible by 3, and we take $k_y=(L/3)-k_x$, and 
$k_z=(L/3)+k_x$, then $\omega^2=1.5$ for {\it any} $k_x$.  These 
degeneracies are lattice artifacts.  We conjecture that they are not 
present when $L$ is a prime number, and we have verified that they are 
absent when $L$= 11, 31 or 53.  When removing degeneracy is an important
consideration, we will choose $L$ to be a prime number. 

Further degeneracies are associated with symmetries of the cubic lattice.
The factor $\omega (k_x,k_y,k_z)$ is unchanged when the momentum components 
are permuted (factor of 1, 3 or 6) or $k_\mu$ is replaced by $L-k_\mu$ 
(factor of 8).  There are two transverse polarizations, 
so up to 96 tranverse polarization states have the same energy.  The 
degeneracy of transverse pair states is up to 48.  When states are grouped 
to form representations of the rotation group of the cube, these states are 
organized into 24 orthogonal linear combinations of differing ``spin''.  
This leaves a two-fold degeneracy, and we expect the transverse 
polarization spin zero states we study to be doubly degenerate.

We can choose polarization vectors so that the remaining double degeneracy 
is labeled by the polarization index $s=1,2$.  We generate 
states that are representations of the rotation group of the cube by 
specifying a ``seed'' wave vector in the interval
\begin{equation}
\label{eq:seed}
0\le k_x\le k_y\le k_z\le (L-1)/2;\ \ ({\rm exclude}\ {\bf k}=0.)
\end{equation}
Then the sets $\{R{\bf k}\}$ and $\{{\bf L-}R{\bf k}\}$ span all $L^3-1$ 
wave vectors as $R$ ranges over the  24 elements of the rotation group and 
$\bf k$ ranges over the set of seed vectors.  For vectors in 
Eq. (\ref{eq:seed}) we choose $\epsilon^{(1)}_\mu({\bf k})$ so 
that $\epsilon^{(1)}_x({\bf k})=0$ and $\epsilon^{(1)}_y({\bf k})>0$.    
${\bf\epsilon}^{(2)}={\bf\epsilon}^{(3)}{\bf\times\epsilon}^{(1)}$.  Then 
choose 
\begin{equation}
\epsilon^{(s)}(R{\bf k})=\epsilon^{(s)}({\bf L-}R{\bf k})
=R\epsilon^{(s)}({\bf k}).
\end{equation}  
These choices assure that all states contributing to a 
representation of the rotation group have the same polarization quantum 
number.  The index $s$ becomes a quantum number that distinguishes the 
two degenerate spin 0 states at each energy eigenvalue.

The results we expect for pure SU(3) differ from these.  The 
operators $Q_a({\bf s})$ satisfy the Lie algebra of SU(3), so the spectrum 
of $Q^2$ is discrete, consisting of sums of integer multiples of the 
quadratic Casimir eigenvalues of SU(3): $\lambda(p,q)=$ 4/3, 8/3, $\dots$.   
Thus $g^2Q^2$, which appears here, has a discrete spectrum, but one whose 
spacing goes to zero at $g=0$.  The disappearance of the gaps between 
states rationalizes the continuous spectrum we have found for $g^2Q^2$ 
when we use the weak coupling Hamiltonian.  Note that in the weak coupling 
limit the eight gauge generators at each site commute, so the local gauge 
invariance group becomes Abelian, and is no longer SU(3).

\section{\label{sec:level3}SU(3) Hamiltonian Using Our Basis}

The basis states we use are Fock states generated by applying products of 
operators $E_a^{(s)\dagger} ({\bf k})$ to the ground state of generator 
$G_B$, Eq. (\ref{eq:generator}).  We write Hamiltonian $H_A$, 
Eq. (\ref{eq:ham}), as an operator on this basis, using the 
matrix elements of Ref.~\cite{E}.  The correspondence between the Fock 
states and the states in Ref.~\cite{E} is
\begin{equation}
|\phi\rangle\leftrightarrow|0\rangle\sqrt{S_1},\qquad|\phi_a\rangle
\leftrightarrow -iA^{\dagger}_a|0\rangle\sqrt{S_2}.
\end{equation}

For example, the four matrix elements among $|\phi\rangle$ and 
$|\phi_a\rangle$ are correctly given by the operators 
\begin{equation}
\label{eq:electric}
{\cal J}_{L(R),a}=-i\sqrt{\frac{S_2}{S_1}}\left(A_a-A^{\dagger}_a\right)
+1(-1)\frac{i}{2}f_{pqa}A_p^{\dagger}A_q,
\end{equation}
where repeated color indices are summed. 

In the Hamiltonian we also need the operator ${\cal J}^2$.  Matrix elements 
of this operator are given in Ref.~\cite{E}, but we can also use the 
relations ${\cal J}^2={\cal J}_{L,a}{\cal J}_{L,a}
={\cal J}_{R,a}{\cal J}_{R,a}$ and Eq. (\ref{eq:electric}) to compute 
the operator.  The expressions differ because when we begin with 
Eq. (\ref{eq:electric}), intermediate states between the two factors 
do not form a complete set.  A choice must be made for ${\cal J}^2$, 
and we will use the square of the operator in Eq. (\ref{eq:electric}) 
in this paper because that is what appears naturally in the operator 
$Q^2=Q_aQ_a$.
\begin{equation}
\label{eq:electric1}
{\cal J}^2=-\frac{S_2}{S_1}\left(A_a-A^{\dagger}_a\right)^2+\frac{3}{4}
A^{\dagger}_aA_a-\frac{1}{4}f_{pqa}f_{rsa}A^{\dagger}_pA^{\dagger}_r
A_qA_s.
\end{equation}

The operator reproducing the four matrix elements of $U$ among 
$|\phi\rangle$ and $|\phi_a\rangle$ is
\begin{eqnarray}
\label{eq:ewe}
U&=&\frac{K_0}{S_1}-\frac{iK_0}{4\sqrt{S_1S_2}}\lambda_c(A_c
+A_c^{\dagger})\\
& &+\left[\frac{DK_0}{2S_2}-\frac{K_0}{S_1}\right]A_c^{\dagger}A_c
+\frac{K_3}{2S_2}\left\{\lambda_c,\lambda_d\right\}A_c^{\dagger}A_d
\nonumber
\end{eqnarray}
The matrix element of the operator $U^{\dagger}$ on our basis is obtained 
by changing the sign of the second term in this equation.
Expressions for the matrix element coefficients from Ref.~\cite{E} are:
\begin{eqnarray}
K_0&=&\frac{1}{3}\sum_{p,q}e(p,q)f(p,q),\\
K_3&=&\frac{1}{4}\sum_{p,q}e(p,q)f(p,q)\left[-\frac{\gamma(p,q)}{20}+
\frac{\gamma(q,p)}{20}-\frac{1}{9}\right],\nonumber\\
DK_0&=&\frac{1}{3}\sum_{p,q}e(p,q)f(p,q)\nonumber\\
& &\times\left[\frac{\gamma(p,q)}{8}+\frac{\gamma(q,p)}{40}
-\frac{\lambda(p,q)}{5}-\frac{1}{18}\right],\nonumber\\
e(p,q)&=&d(p,q)e^{-t\lambda(p,q)},\nonumber\\
f(p,q)&=&e(p+1,q)+e(p,q-1)+e(p-1,q+1),\nonumber\\
\gamma(p,q)&=&(2p+q)(p+2q+6_p-q+3)/9.\nonumber
\end{eqnarray}
$\gamma$ is the eigenvalue of the {\it cubic} Casimir operator of SU(3).
When these operators are inserted into Hamiltonian terms of the type 
Tr$(U_1U_2U^{\dagger}_3U^{\dagger}_4)$, expressions involving up to 
eight $A$'s and $A^\dagger$'s appear.  The variational Hamiltonian matrix 
is generated by taking matrix elements of these expressions in our basis.

Simplifications occur at weak coupling, which we consider at this point.  
An imaportant test of the adequacy of our basis is to show that the 
Kogut-Susskind Hamiltonian assumes the form $H_W$ at weak coupling.  
At weak coupling, the diffusion time  $t=g^2\tau$ is small because 
we have seen that $\tau$ remains finite as $g$ is taken to zero.  Many terms 
contribute to the sums for $S_1$ and $S_2$ when $t$ is small, and the 
leading behavior at small $t$ is 
$\sqrt{S_2/S_1}\sim 1/(2\sqrt{t})=1/(2g\sqrt{\tau})$.  Thus, at weak 
coupling the non-Abelian terms in ${\cal J}_{L,R}$ can be ignored.  We 
use Eq. (\ref{eq:weakops2}), now with the choice $\sigma=\tau$.  Then at 
weak coupling, we obtain a representation identical to that in 
Sec.~\ref{sec:level2}:
\begin{equation}
{\cal J}_{L,a}={\cal J}_{R,a}=\frac{1}{ig}\frac{\partial}{\partial R_a}.
\end{equation}

At small $t$ the coefficients in $U$ have the behaviors
\begin{eqnarray}
\label{eq:lim1}
\frac{K_0}{S_1}=1-2t/3,\\
\label{eq:lim2}
\frac{K_0}{4\sqrt{S_1S_2}}=\frac{\sqrt{t}}{2},\\
\label{eq:lim3}
\left[\frac{DK_0}{2S_2}-\frac{K_0}{S_1}\right]=\frac{25 t^2}{12},\\
\label{eq:lim4}
\frac{K_3}{2S_2}=-\frac{t}{8}.
\end{eqnarray}
We retain contributions to ${\rm Tr}\left(U_1U_2U^{\dagger}_3
U^{\dagger}_4\right)$ that are order 
$t$ or larger.  Therefore the term in Eq. (\ref{eq:lim3}) can be ignored.  
Other contributions are:
\begin{eqnarray}
& &{\rm Tr}\left[\left(\frac {K_0}{S_1}\right)^4\right]=3-8t,\\
& &{\rm Tr}\left[\left(\frac {K_0}{S_1}\right)^2\left(
\frac{-iK_0}{4\sqrt{S_1S_2}}
\right)^2\lambda_b\lambda_c\right]\left(A_{1,b}+A^{\dagger}_{1,b}\right)
\nonumber\\
& &\times\left(A_{2,c}+A^{\dagger}_{2,c}\right)=-\frac {t}{4\tau}(2R_1R_2),
\nonumber\\
& &{\rm Tr}\left[\left(\frac {K_0}{S_1}\right)^3\left(\frac{K_3}{2S_2}
\right)\{\lambda_b,\lambda_c\}\right]A^{\dagger}_{1,b}A_{1,c}\nonumber\\
& &=-\frac{t}{4}(2A^{\dagger}_{1,b}A_{1,b})
=-\frac{t}{4}(A^{\dagger}_{1,b}A_{1,b}+A_{1,b}A^{\dagger}_{1,b}-8)
\nonumber\\
& &=-\frac{t}{4\tau}R_1^2+\frac{t}{4}\left[(A_{1,b})^2+
(A^{\dagger}_{1,b})^2\right]+2t.\nonumber
\end{eqnarray} 
Using these expressions and $t/\tau=g^2$, we have 
\begin{eqnarray}
\label{eq:ham3}
H_A& &=-\frac{1}{2}\sum_{{\bf s},m,a}\frac{\partial^2}{\partial 
R_a({\bf s},m)^2}\\
& &+\frac{1}{2}\sum_{{\bf s},m<n,a}[R_a({\bf s},m)+
R_a({\bf s+\hat m},n)\nonumber\\
& &-R_a({\bf s+\hat n},m)-R_a({\bf s},n)]^2\nonumber\\
& &+\mu\sum_{{\bf s},m,a}\{[A_a({\bf s},m)]^2+
[A^{\dagger}_a({\bf s},m)]^2\},\nonumber
\end{eqnarray}
where $\mu=-2\tau$.

$H_A$ takes the form it does because of delicate combinations of the small 
$t$ expressions given in Eqs. (\ref{eq:lim1}) - (\ref{eq:lim4}).  The 
coefficient 1/2 of the middle term of Eq. (\ref{eq:ham3}) comes from 
combining Eqs. (\ref{eq:lim2}) and (\ref{eq:lim4}).  There is no additive 
c-number in the Hamiltonian because there is a cancellation between the term 
$-2t/3$ in Eq. (\ref{eq:lim1}) and the term $-t/8$ in Eq. (\ref{eq:lim4}), 
together with the normal ordering constant 8, the number of SU(3) 
generators.  Note that it is surprising that the coefficients we find are 
even rational, because they are ratios of sums that are irrational in their 
small-$t$ limits.  For example, the small-$t$ form of $S_1$ is 
$(\pi\sqrt{3})/(16t^4)$. 

The value $\mu=-2\tau$ is unreliable because we have not included doubly 
excited color/link operators in Eq. (\ref{eq:ewe}).  The reason such 
operators appear in Eq. (\ref{eq:ham3}) (with incorrect coefficient) is that 
the Hamiltonian involves {\it products} of the basic operators.  For these 
terms to contribute to matrix elements, one or more of the states must 
be ``overexcited''.  For this reason it might be thought that it is harmless 
to use the value $\mu=-2\tau$.  That is not the case, however, because the 
Hamiltonian possesses a ground state only for $0\le\mu < 1/(4\tau)$.

The most reasonable choice for $\mu$ is $\mu=0$ because analogous leftover 
terms like $A_a({\bf s_1})A_a({\bf s_2})$, ${\bf s_1}\ne{\bf s_2}$ do not 
appear.  The coefficient of these terms is completely determined by 
Eq. (\ref{eq:ewe}).  

At this point $H_A$, Eq. (\ref{eq:ham3}), is identical to $H_W$, 
Eq. (\ref{eq:osc}).  The expressions were derived in completely different 
ways, however.  $H_W$ was derived by using weak limit expressions for the 
SU(3) {\it operators}.  When we derived $H_A$, the operators were 
unmodified.  We constructed the Hamiltonian to give correct matrix 
elements on a limited {\it basis} of 256 states per link, a procedure that 
can be employed at any coupling.  It was only when we took $g=0$ at the 
end, that we found $H_A=H_W$.  We also use $H_W$ and $H_A$ differently.  
We simply diagonalized $H_W$ to obtain exact weak coupling results.  $H_A$ 
will be evaluated on our basis of states having small $N_{ave}$ and therefore 
minimal contribution from overexcited states for each color/link degree of 
freedom. Our demonstration that the weak coupling form of $H_A$ coincides 
with $H_W$ is therefore only a first necessary check on the viability of 
the use of our basis.  What remains to be demonstrated is that the 
spectrum of gauge-invariant states we obtain (by extrapolation to 
$g^2Q_{ave}^2=0$) is a reasonable approximation to the gauge-invariant 
spectrum of $H_W$.

At weak coupling, $Q^2$ is given by Eq. (\ref{eq:queue}) (with 
$\sigma\rightarrow\tau$), and in momentum space $N$ is
\begin{equation}
N=\sum_{{\bf k},a}\sum_{s=1}^3B_a^{(s)\dagger}({\bf k})B_a^{(s)}
({\bf k}).
\end{equation}
We can now compute $G_B$ in momentum space.  The operators $N$ and 
$g^2Q^2_0$ provide a 
restoring term for the longitudinal mode, so for $G_B$ we can 
perform a Bogoliubov transformation for both longitudinal and transverse 
modes.  The parameters of the Bogoliubov transformations are
\begin{eqnarray}
4\theta^{(T)}({\bf k})&=&-\ln\left\{\frac{16\tau^2[\omega^2({\bf k})
+\xi/8\tau]}{1+2\xi\tau}\right\},\nonumber\\
4\theta^{(3)}({\bf k})&=&\ln\left\{\frac{8\eta\omega^2({\bf k})+1
+2\xi\tau}{2\xi\tau}\right\}.
\end{eqnarray}
Then
\begin{eqnarray}
G_B&=&G_{B,0}+\sum_{{\bf k},a}\sum_{s=1}^2G_{B,1,T}({\bf k})
E^{(s)\dagger}_a({\bf k})E^{(s)}_a({\bf k})\nonumber\\
& &+\sum_{{\bf k},a}G_{B,1,L}({\bf k})E^{(3)\dagger}_a({\bf k})
E^{(s)}_a({\bf k}),
\end{eqnarray}
where
\begin{eqnarray}
G_{B,0}/L^3&=&16\sqrt{1+2\xi\tau}T_1(\alpha)\\
& &+8\sqrt{\xi\eta/\tau}T_1(\beta)-12\xi,\nonumber\\
G_{B,1,T}&=&2\sqrt{[1+2\xi\tau][\omega^2({\bf k})+\alpha]},
\nonumber\\
G_{B,1,L}&=&2\sqrt{[\xi\eta/\tau][\omega^2({\bf k})+\beta]}.
\nonumber
\end{eqnarray}
The new expressions in these formulas are:
\begin{eqnarray}
\alpha&=&\xi/8\tau,\quad\beta=(1+2\xi\tau)/8\eta,\\
T_{1(2)}(\gamma)&=&\frac{1}{L^3}\sum_{\bf k}\left[\omega^2({\bf k})
+\gamma\right]^{1/2(-1/2)}.\nonumber
\end{eqnarray}
The average quadratic Casimir operator is:
\begin{equation}
g^2Q_{ave}^2=4\sqrt{\frac{\xi}{\tau\eta}}\left[T_1(\beta)
-\beta T_2(\beta)\right].
\end{equation}
We also have:
\begin{eqnarray}
N&=&N_0+\sum_{{\bf k},a}\sum_{s=1}^2\left\{N_{1,T}({\bf k})
E^{(s)\dagger}_a({\bf k})E^{(s)}_a({\bf k})\right.\nonumber\\
& &+N_{2,T}({\bf k})\left[E^{(s)}_a({\bf k})E^{(s)}_a({\bf L-k})
\right.\nonumber\\
& &\left.+E^{(s)\dagger}_a({\bf k})E^{(s)\dagger}_a({\bf L-k})
\right]\nonumber\\
& &+\sum_{{\bf k},a}\left\{N_{1,L}({\bf k})
E^{(3)\dagger}_a({\bf k})E^{(3)}_a({\bf k})\right.\nonumber\\
& &+N_{2,L}({\bf k})\left[E^{(3)}_a({\bf k})E^{(3)}_a({\bf L-k})
\right.\nonumber\\
& &\left.\left.+E^{(3)\dagger}_a({\bf k})E^{(3)\dagger}_a({\bf L-k})
\right]\right\},\\
\frac{N_0}{L^3}&=&\frac{16\tau}{\sqrt{1+2\xi\tau}}T_1(\alpha)
+\frac{\sqrt{1+2\xi\tau}}{\tau}T_2(\alpha)
\nonumber\\
& &+4\sqrt{\frac{\eta}{\xi\tau}}T_1(\beta)+\sqrt{\frac{\xi\tau}
{\eta}}T_2(\beta)-12,\nonumber\\
N_{1,T}&=&\frac{2\tau\sqrt{\omega^2({\bf k})+\alpha}}{\sqrt{1+2\xi\tau}}
+\frac{\sqrt{1+2\xi\tau}}{8\tau\sqrt{\omega^2({\bf k})+\alpha}},\nonumber
\\
N_{2,T}&=&\frac{\tau\sqrt{\omega^2({\bf k})+\alpha}}{\sqrt{1+2\xi\tau}}
-\frac{\sqrt{1+2\xi\tau}}{16\tau\sqrt{\omega^2({\bf k})+\alpha}},\nonumber
\\
N_{1,L}&=&\sqrt{\frac{\eta}{\xi\tau}}\sqrt{\omega^2({\bf k})
+\beta}+\frac{1}{4}\sqrt{\frac{\xi\tau}{\eta}}\frac{1}{\sqrt
{\omega^2({\bf k})+\beta}},\nonumber\\
N_{2,L}&=&-\frac{1}{2}\sqrt{\frac{\eta}{\xi\tau}}\sqrt{\omega^2({\bf k})
+\beta}+\frac{1}{8}\sqrt{\frac{\xi\tau}{\eta}}\frac{1}{\sqrt
{\omega^2({\bf k})+\beta}}.\nonumber
\end{eqnarray}

The mean excitation per color/link degree of freedom is 
\begin{eqnarray}
\label{eq:ave}
N_{ave}&=&\frac{N_0}{24L^3}=\frac{2\tau}{3\sqrt{1+2\xi\tau}}T_1(\alpha)
+\frac{\sqrt{1+2\xi\tau}}{24\tau}T_2(\alpha)
\nonumber\\
& &+\frac{1}{6}\sqrt{\frac{\eta}{\xi\tau}}T_1(\beta)+\frac{1}{24}
\sqrt{\frac{\xi\tau}
{\eta}}T_2(\beta)-\frac{1}{2}.
\end{eqnarray}
It can be shown that the probability that a color/link degree of freedom is 
``overexcited'' is
\begin{equation}
P=1-\frac{2N_{ave}^2-2J^2+3N_{ave}+1}{[(1+N_{ave})^2-J^2]^{3/2}},
\end{equation}
where
\begin{eqnarray}
J&=&=-\frac{2\tau}{3\sqrt{1+2\xi\tau}}T_1(\alpha)
+\frac{\sqrt{1+2\xi\tau}}{24\tau}T_2(\alpha)\\
& &+\frac{1}{6}\sqrt{\frac{\eta}{\xi\tau}}T_1(\beta)-\frac{1}{24}
\sqrt{\frac{\xi\tau}{\eta}}T_2(\beta).\nonumber
\end{eqnarray}

At weak coupling, $H_{A,1}=g^2Q_1^2=0$.  Computationally, it is convenient 
to regroup the terms that remain:
\begin{equation}
\label{eq:finalham1}
H_B=G_C-\xi N',
\end{equation}
where
\begin{eqnarray}
G_C&=&[G_{B,0}-\xi N_0]+\sum_{{\bf k},a}\sum_{s=1}^2[G_{B,1,T}({\bf k})
-\xi N_{1,T}({\bf k})]\nonumber\\
& &\times E^{(s)\dagger}_a({\bf k})E^{(s)}_a({\bf k})\nonumber\\
& &+\sum_{{\bf k},a}[G_{B,1,L}({\bf k})-\xi N_{1,L}({\bf k})]\\
& &\times E^{(3)\dagger}_a({\bf k})E^{(3)}_a({\bf k}).\nonumber
\end{eqnarray}
$G_C$ incorporates the ``diagonal'' terms of $-\xi N$, and has the same 
eigenstates as the generator $G_B$.  The remaining term is
\begin{eqnarray}
&-\xi N'&=-\xi\sum_{{\bf k},a}\sum_{s=1}^2 N_{2,T}({\bf k})\left[
E^{(s)}_a({\bf k})E^{(s)}_a({\bf L-k})\right.\nonumber\\
& &\left.+E^{(s)\dagger}_a({\bf k})E^{(s)\dagger}_a({\bf L-k})\right]
\\
& &-\xi\sum_{{\bf k},a}N_{2,L}({\bf k})\left[E^{(3)}_a({\bf k})
E^{(3)}_a({\bf L-k})\right.\nonumber\\
& &\left.+E^{(3)\dagger}_a({\bf k})E^{(3)\dagger}_a({\bf L-k})
\right].\nonumber
\end{eqnarray}

At general coupling, there are many terms in $H_B$ that are missing in 
its weak coupling limit.  We will discuss this case later.

\section{\label{sec:level4}The Basis and its Properties}

Generator $G_C$ has an infinite number of eigenstates, and we can 
accommodate only a finite number of these in our basis.  The goal is a 
basis that allows us to determine the spectrum of low energy, 
gauge-invariant eigenstates of $H_B$, so we include only the low energy 
spin zero eigenstates of $G_C$.  The lowest energy such states are spin 
zero gluon pair states of the form
\begin{equation}
|{\bf k},s\rangle=A({\bf k})\sum_{R,a}E^{(s)\dagger}_a(R{\bf k})
E^{(s)\dagger}_a(R[{\bf L-k}])|0\rangle.
\end{equation}
The wave vector here is one of the ``seed'' vectors of 
Eq. (\ref{eq:seed}).  The number of such seed vectors is 
$-1+(L+1)(L+3)(L+5)/48$.  For example, when $L=31$, there are 815 single 
pair states of each polarization.  These states are composed of two gluons 
and thus are charge conjugation even.  Both gluons have the same spin index 
so the states have even parity.  They constitute a basis for computing the 
spectrum of $0^{++}$ eigenstates. 

The normalization constants depend on the seed momentum because as 
$R$ ranges over the 24 group elements, the number of independent states 
that appear can be 3, 4, 6, 12 or 24, depending on $\bf k$.  Denote 
the number of independent states by $V$; then $A({\bf k})=\sqrt{2V}/96$.

When $\eta$ is reasonably small (say, 0.1), the single longitudinal pair 
states are closely spaced and lie in an isolated band.  (At $\eta=0$ the 
band width is zero.)  There is then a gap between 
these single longitudinal pair states and the lowest energy states composed 
of two longitudinal pairs.  So, when $\eta$ is small, our basis will 
consist of the single longitudinal pair states and those single transverse 
pair states that have energies within the band described above.  In general, 
only a small number of single transverse pair states falls within the band of 
single longitudinal pair states.  This is not a serious limitation because 
in lattice gauge theory the objective is always to determine the first few 
gauge invariant states of given spin, charge conjugation and parity.

As $\eta$ is increased, the content of the basis changes.  Then the spacing 
of the single longitudinal pair states increases, and further single 
transverse pair states fall within the band. Eventually, the highest energy 
single longitudinal pair state has greater energy than the lowest two 
longitudinal pair state.  When that happens, we take the basis 
to include only those single pair states (longitudinal and transverse) 
that lie below the two longitudinal pair threshold.  Note that we do not 
want to increase $\eta$ too much anyway, because for $\eta >1$, 
$\eta g^2Q^2$ becomes the dominant operator in the generator. 

When we extrapolate to $g^2Q^2_{ave}=0$, the longitudinal polarization 
pair states in the basis behave quite differently than the transverse 
pair states.  The longitudinal states all extrapolate to energy zero, 
where they merge with the ground state.  Then they become part of the 
continuum of longitudinal states found in the exact weak coupling solution.  
In our basis, these states are promoted to finite energy by the additional 
operators in $G_C$.  

It would be pleasant to simply omit the longitudinal polarization pairs 
in our basis.   However, the interaction term $-\xi N'$ connects them to 
the basis vacuum state $|0\rangle$, which is included in the basis.  The 
longitudinal pairs and transverse pairs became coupled when we introduced the 
operator $N$ to suppress the ``overexcited'' states introduced by our use of 
boson operators. 

The transverse pair states have the same expectation of $Q^2$ as the 
basis ground state.  That is, they are excitations of energy only.  When 
we extrapolate to $g^2Q_{ave}^2=0$ the transverse states approach finite 
limits, and they become gauge invariant: $Q^2|\psi\rangle=0$. We will see 
that these energy limits approximate the energies we found for the gauge 
invariant pair states of Sec.~\ref{sec:level2}.  
  
\subsection{The Ground State Energy Density}

The energy density we study is the expectation value 
${\cal E}=\langle G_C\rangle/L^3$, as a function of $g^2Q_{ave}^2$.  
We study this quantity rather than $\langle G_B\rangle/L^3$ because we want 
to compare our results with the weak coupling expression, 
Eq. (\ref{eq:gsenergy}).  This ground state energy density has no physical 
importance, but an accurate approximation scheme should make a reasonable 
prediction.  We have
\begin{eqnarray}
{\cal E}&=&\frac{16(1+\xi\tau)}{\sqrt{1+2\xi\tau}}T_1(\alpha)\\
& &-\frac{\xi}{\tau}\sqrt{1+2\xi\tau}T_2(\alpha)+\frac{1}{2}
\sqrt{\frac{\xi}{\eta\tau}}T_2(\beta).\nonumber
\end{eqnarray}
\begin{figure}
\includegraphics[angle=-90,width=8cm,keepaspectratio=true,
scale=0.5]{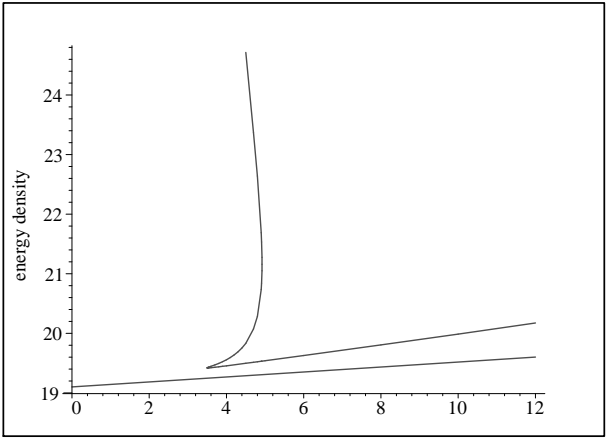}
\caption{\label{fig:fig1}Ground state energy density vs. $g^2Q_{ave}^2$.  
$L=31$, and $\eta=0.1$. The lower line is the ground state energy density 
of $H_W$, Eq. (\ref{eq:gsenergy}).}
\end{figure}
In Fig.~\ref{fig:fig1} we show this ground state energy density when 
$L=31$ and $\eta=0.1$.  The parameters $\xi$ and $\tau$ are chosen so 
$N_{ave}=0.5$, and $g^2Q^2_{ave}$ takes a value on the abscissa of 
Fig.~\ref{fig:fig1}.  The lower branch of the cusped curve is the ground 
state energy density produced by our basis.  To obtain the energy density at 
the gauge invariant limit, we fit a quadratic curve to the lower branch of 
the cusped curve, requiring that the curves intersect at the points 
$g^2Q_{ave}^2=$5.02, 5.00 and 4.98.  The energy density on this curve at 
$g^2Q_{ave}^2=0$ is our extrapolant: ${\cal E}=19.1192$, which is to be 
compared with the true ground state energy density ${\cal E}_0$=19.1008.  
The fractional error is 0.000974.  Very similar results are obtained when 
the quadratic curve is fitted to the lower branch of the cusped curve at 
other points.  The probability of ``overexcitation'' of 
a color/link degree of freedom is in the range 0.11 to 0.12 for the ground 
states parameters that produce the lower branch of the cusped curve.

\begin{figure}
\includegraphics[angle=-90,width=8cm,keepaspectratio=true,
scale=0.5]{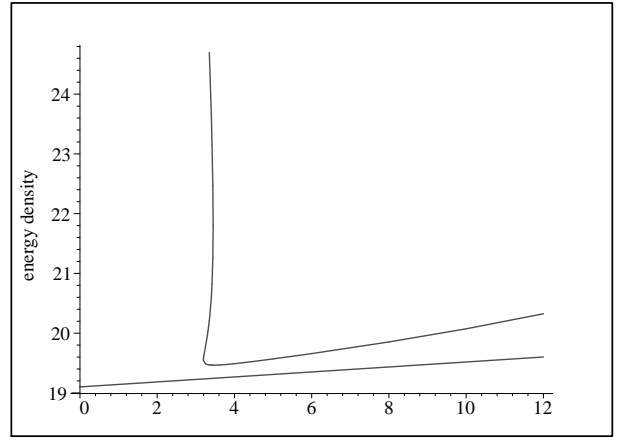}
\caption{\label{fig:fig2}Ground state energy density vs. $g^2Q_{ave}^2$.  
$L=51$, and $\eta=0.5$.}
\end{figure}

Fig.~\ref{fig:fig2} shows the ground state energy density when $L$ and 
$\eta$ are larger.  The extrapolated and true ground state energy densities 
are ${\cal E}=19.2356$ and ${\cal E}_0=19.1008$ for a fractional error 
0.00706.

The ground state energy curve when $\eta=.01$ resembles 
Fig.~\ref{fig:fig1}, but the cusp is more pronounced.  In this case, the 
fractional error of the extrapolant is 0.000316 when $L=31$.  We conclude 
that the results for the ground state are stable when $\eta$ is scaled by 
a factor of 50.

\subsection{The Basis and Excited States}

The numerical data for the excited basis states is most easily explained 
by laying out a typical example.  Consider the basis when $L=31$ and 
$\eta=0.1$.  We take $\tau=0.428886$ and $\xi=0.0567920$, which sets 
$N_{ave}$=0.5 and $g^2Q^2_{ave}=4.000$, on the lower branch of 
Fig.~\ref{fig:fig1}.  The basis consists of 815 single longitudinal pair 
states, and one single transverse pair state.  The transverse state 
energy is independent polarization $s$, so in fact it is doubly degenerate, 
corresponding to the two-fold degeneracy we found for the spin zero 
eigenstates of $H_W$.  These states all lie in the energy interval 
$0.252337\le E\le 0.470661$ above the ground state.  Our convention is 
to order the states with increasing energy, with the ground state assigned 
ordinal number zero.  Then when $g^2Q^2_{ave}=4.000$, the transverse pair 
state has ordinal number 777, and its  energy is 0.447724.

\begin{figure}
\includegraphics[angle=-90,width=8cm,keepaspectratio=true,
scale=0.5]{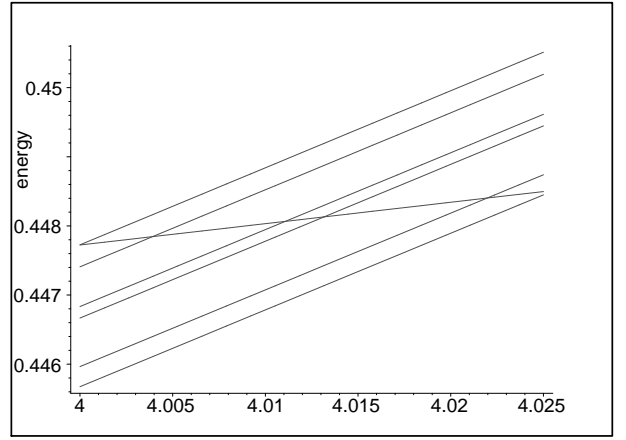}
\caption{\label{fig:fig3}Excited state energy (above ground state) vs. 
$g^2Q_{ave}^2$, for eigenstates of $G_C$.  Ordinal states 772 to 778 are 
shown.} 
\end{figure}

The basis energies vary with $g^2Q^2$, and Fig.~\ref{fig:fig3} shows seven 
states near $g^2Q_{ave}^2=4.000$.  The longitudinal pair states never cross 
each 
other, but the transverse pair state crosses four longitudinal pair states 
in the region depicted.  The level crossings are consistent with the 
fact that the longitudinal pair energies and the transverse pair energy are 
directed toward different limits as $g^2Q_{ave}^2$ is reduced.

We next repeat the computations of the energies when 
$g^2Q^2_{ave}=4.000\pm 0.0200$, fit quadratic curves through the three 
data points and extrapolate to $g^2Q^2_{ave}=0$.  The extrapolated energies 
of all 815 longitudinal pair states are very close to zero.  For example, 
ordinal state 100 has energy 0.313099 at $g^2Q^2_{ave}=4.000$.  The 
extrapolated energy is -0.0110660.  To merge with the ground state, the 
extrapolated energy would have to be zero.  The fractional error is about 
-0.035.  As the ordinal quantum number of the longitudinal pair increases, 
the extrapolated energy of the longitudinal pair also increases.  The 
range over all 815 longitudinal pairs is from -0.023 to +0.017.  Thus, to 
accuracy we might expect, the longitudinal states merge with the ground 
state at $g^2Q^2_{ave}=0$.  For the transverse state, however, the 
extrapolated energy is 0.3788, which should be compared to the exact 
result 0.4047 (from $H_W$) on this lattice.
\begin{table}
\caption{\label{tab:table1}  Extrapolated first excited tranverse pair 
energy.  Exact energy is 0.4047.  For each value of $g^2Q^2_{ave}$ the top 
entry occurs when the transverse pair is near the top of the longitudinal 
pair basis, as indicated by the ordinal number.}
\begin{ruledtabular}
\begin{tabular}{cccc}
$g^2Q^2_{ave}$:&4.000&5.000&7.000\\
$\eta$&&&\\
\hline
0.043&--&--&0.3552\footnotemark[1]\\
0.070&--&0.3843\footnotemark[2]&0.3379\\
0.095&0.3805\footnotemark[3]&0.3743&0.3236\\
0.100&0.3788&0.3723&0.3236\\
0.300&0.3076&0.3024&0.2453\\
0.500&0.2483&0.2516&0.2018\\
\end{tabular}
\end{ruledtabular}
\footnotetext[1]{Ordinal number 800/816.}
\footnotetext[2]{Ordinal number 804/816.}
\footnotetext[3]{Ordinal number 804/816.}
\end{table}
\begin{table}
\caption{\label{tab:table2}  Extrapolated second excited tranverse pair 
energy.  Exact energy is 0.5723.}
\begin{ruledtabular}
\begin{tabular}{cccc}
$g^2Q^2_{ave}$:&4.000&5.000&7.000\\
$\eta$&&&\\
\hline
0.070&--&--&0.5490\footnotemark[1]\\
0.100&--&--&0.5357\\
0.115&--&0.5692\footnotemark[2]&0.5259\\
0.190&0.5501\footnotemark[3]&0.5493&0.4900\\
0.300&0.5232&0.5174&0.4456\\
0.500&0.4725&0.4630&0.3848\\
\end{tabular}
\end{ruledtabular}
\footnotetext[1]{Ordinal number 810/817.}
\footnotetext[2]{Ordinal number 793/817.}
\footnotetext[3]{Ordinal number 587/595.}
\end{table}

The most interesting issue is how close the extrapolated transverse pair 
energies approach exact results when we vary $\eta$ and the $g^2Q^2_{ave}$ 
from which we start the extrapolation.  These results are given for the two 
lowest energy transverse pair states in Tables \ref{tab:table1} and 
\ref{tab:table2}.  Some table entries are empty because when $\eta$ is 
sufficiently small, even the lowest transverse pair states lie above the 
band of longitudinal pair states.  That is, they are not in the 
basis as we have defined it.  The top entry in each column is for 
a value of $\eta$ where the transverse pair state falls just below the 
highest energy longitudinal pair states in the the basis, as indicated.  

The next question is how to select a prediction, since the table entries 
vary considerably.  For a given column in the tables we choose the entry 
with the smallest value of $\eta$, since this parameter fixes the strength of 
the term $g^2\eta Q^2_0$ in $G_B$, which perturbs the basis states relative 
to the eigenstates of $H_A$.  For each column, this selects the largest entry 
in the column.  Given that, it is natural to select the largest entry in the 
table as the prediction.  The fractional error of this prediction for 
the first excited pair state is then (0.3843-0.4047)/0.4047=-0.050.  
For the second excited pair state the fractional error is -0.005, and for 
the third excited pair state the error is +0.002.

In the next section we will see that the reliability of these extrapolations 
for transverse pair states is crucial for our approach to pure SU(3) lattice 
gauge theory.

\section{\label{sec:level5}Eigenstates of the Hamiltonian}

The basis consists of one pair eigenstates of $G_C$, to which operator we 
must add the interaction term $=-\xi N'$ to obtain Hamiltonian $H_B$.  On our 
basis, the interaction term is a rank one operator.  It may be written
\begin{eqnarray}
-\xi N'&=&-(384)\xi A({\bf k})\sum_{s=1}^3\sum_{\bf k}
N_{2,s}({\bf k})\\
&\times&\{|{\bf k},s\rangle\langle 0|+|0\rangle\langle{\bf k},s|\}.
\nonumber
\end{eqnarray}
The sums are over the seed momenta only.  

We can immediately write half the transverse eigenstates of $H_B$.  Recall 
that the states $|{\bf k},s\rangle$ with $s=1,2$ are degenerate eigenstates 
of $G_C$.  This means that the orthonormal combinations 
\begin{equation}
|{\bf k},\pm\rangle=\frac{1}{\sqrt{2}}\left\{|{\bf k},1\rangle\pm 
|{\bf k},2\rangle\right\}
\end{equation} 
are also eigenstates with the common eigenvalue.  However, it is evident 
that $(-\xi N')|{\bf k},-\rangle=0$, so the $|{\bf k},-\rangle$ states are 
already eigenstates of $H_B$.  This means the predications we extracted 
from Tables \ref{tab:table1} and \ref{tab:table2} are extrapolated 
eigenvalues of both $G_C$ and $H_B$ for these states.  The remaining task, 
then, is to see whether the states $|0\rangle$, $|{\bf k},+\rangle$ and 
$|{\bf k},L\rangle$, which are connected by operator $-\xi N'$ form gauge 
invariant eigenstates of $H_B$ that extrapolate to the limits in 
Tables \ref{tab:table1} and \ref{tab:table2}.

Because $-\xi N'$ is a rank one operator on our basis, it is easy to 
construct eigenstates of $H_B$ despite the large size of the basis.  
Let $|n\rangle$ denote a pair state belonging to our basis, and let 
$E_n$ be the corresponding eigenvalue of $G_C$.  An eigenstate of 
$H_B$ will be a linear combination of our basis states:
\begin{equation}
|\psi\rangle=\alpha |0\rangle+\sum_n\beta_n|n\rangle.
\end{equation}
Then the eigenvalue equation
\begin{equation}
H_B|\psi\rangle=(G_C-\xi N')|\psi\rangle={\cal E}|\psi\rangle
\end{equation}
leads to the relations 
\begin{eqnarray}
\label{eq:ampl}
\beta_n&=&\frac{\langle n|(-\xi N')|0\rangle}{{\cal E}-E_n}\alpha,\\
{\cal E}\alpha&=&\sum_n\langle0|(-\xi N')|n\rangle\beta_n.
\end{eqnarray}
Substituting the first expression into the second, we obtain the 
eigenvalue equation
\begin{equation}
\label{eq:eigenvalues}
{\cal E}=\sum_n\frac{|\langle 0|(-\xi N')|n\rangle|^2}{{\cal E}-E_n}.
\end{equation}

It is easy to locate the solutions of the eigenvalue equation.  Suppose 
the basis eigenvalues $E_n$ are labeled so they increase monotonically 
with $n$.  
Then as ${\cal E}$ increases through the interval $[E_n,E_{n+1}]$, the 
right side of Eqn. (\ref{eq:eigenvalues}) decreases monotonically from 
$+\infty$ to $-\infty$.  As ${\cal E}$ varies in this way, the left side 
increases monotonically.  Thus, there is one solution of the eigenvalue 
equation in the interval.  In addition to these solutions, there is one 
solution below ${\cal E}=0$, the new ground state, and one solution above 
the largest $E_n$.  

\begin{figure}
\includegraphics[angle=-90,width=8cm,keepaspectratio=true,
scale=0.5]{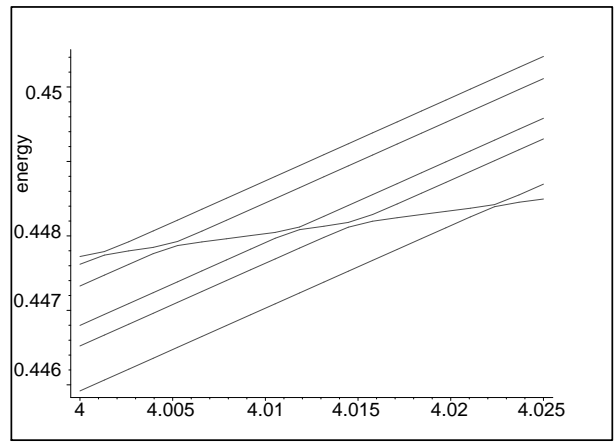}
\caption{\label{fig:fig4}Excited state energy (above ground state) vs. 
$g^2Q_{ave}^2$, for eigenstates of $H_B$ Ordinal states 772 to 777 are 
shown.} 
\end{figure}

Fig.~\ref{fig:fig4} shows the eigenstates of $H_B$ in the region 
of the spectrum displayed in Fig.~\ref{fig:fig3}, again on a lattice with 
$L=31$, $\eta=0.1$.  In keeping with the quantum mechanical result that 
interacting states do not cross, states approach each other closely, but 
never cross.  For this reason, there can no longer be a distinct transverse 
polarization state.  However, there is still a trace of the transverse 
basis state.  By comparing Figs.~\ref{fig:fig3} and \ref{fig:fig4}, we can 
see how trajectories of eigenstates of $H_B$ bend to follow the path of the 
transverse basis state.  Two observations explain why this happens.  First, 
when a transverse basis state crosses a longitudinal basis state, there must 
be an eigenstate of $H_B$ with ${\cal E}=E_L=E_T$.  (The eigenvector is a 
linear combination of the two crossing basis states.)  Between such 
intersections, eigenvalue equation (\ref{eq:eigenvalues})  still has a 
solution close the $E_T$ because in this regime, 
$|\langle 0|(-\xi N')|T\rangle|^2|/\langle 0|(-\xi N')|L\rangle|^2\sim 0.1$.

The eigenvalues of $H_B$ extrapolate to limits very close to zero, and 
all but two of them behave like nearby eigenvalues of $G_C$.  We mentioned 
that ordinal eigenvalue 100 of $G_C$ extrapolates $0.313099\rightarrow 
-0.0110660$.  Ordinal eigenvalue 100 of $H_B$ extrapolates 
$0.313132\rightarrow -0.0110627$.  

However, two eigenvalues are bizarre.  At $g^2=4.000$, the new ground 
state is ${\cal E}_0=-64.9678$.  Note that the interaction $-\xi N'$ is a 
traceless operator on our basis, so the sum of eigenvalues $\cal E$ is the 
same as the sum of the $E_n$.  This suggests that a very positive eigenvalue 
is to be found.  Sure enough, ${\cal E}_{816}=+65.3527$.  The next lower 
eigenvalue is ${\cal E}_{815}=0.47064$, sandwiched between $E_{815}$ and 
$E_{816}$.

The very positive eigenvalue is an artifact of the cutoff of the basis.  For 
example, if we omit basis state 816, then it is ${\cal E}_{815}$ 
that is very large.  Because of the tracelessness of the interaction, the 
cutoff of the basis also induces the very negative ${\cal E}_0$.  

This pathology largely disappears when we extrapolate.  For ${\cal E}_0$, 
the extrapolation is $-64.9678\rightarrow 0.134561$; for ${\cal E}_{816}$, 
the extrapolation is $+65.3527\rightarrow -0.137485$. The fractional 
deviation of the extrapolated ground state energy from zero is +0.002.
We will take zero to be the extrapolated energy for all these energy 
levels.  

But if all energy levels extrapolate to zero within expected accuracy, 
what became of the gauge invariant basis state that extrapolated 
near energy 0.4047?  At this point, the evidence we have of its presence is 
that eigenstates of $H_B$ take turns following the trajectory of the 
transverse basis state.

This picture is clarified when we follow the ``locus of gauge invariance,'' 
as we vary $g^2Q^2_{ave}$.  The expectation value of the operator $g^2Q^2$ 
in the eigenstates of $G_C$ or $H_B$ may be expressed 
\begin{equation}
\langle |g^2Q^2|\rangle =L^3g^2Q^2_{ave}+\Delta.
\end{equation}
A gauge invariant state results when we take the limit $g^2Q_{ave}
\rightarrow 0$ and find $\Delta =0$ in the limit.  

Consider $\Delta$ for the basis states, the eigenstates of $G_C$.  
A transverse basis state has $\Delta=0$ at all values of 
$q^2Q^2_{ave}$, and so when we extrapolate this variable to zero, the 
transverse basis states become gauge invariant.  The basis states 
$|{\bf k},-\rangle$, share this property, and we have seen that they are 
eigenstates of $H_B$ as well as $G_C$.  

Among the other eigenstates of $H_B$, however, no one state has $\Delta=0$, 
at nonzero $g^2Q^2_{ave}$  For these states, $\Delta$ varies with 
$g^2Q^2_{ave}$.  We find there is an abrupt dip in $\Delta$ where the state 
follows the transverse basis state, that is, when $g^2Q^2_{ave}$ is adjusted 
so ${\cal E}\sim E_T$.  This dip is explained by Eq. (\ref{eq:ampl}), 
which shows that the amplitude of the transverse state in the eigenstate is 
particularly large in this region of $g^2Q^2_{ave}$.
  
\begin{figure}
\includegraphics[angle=-90,width=8cm,keepaspectratio=true,
scale=0.5]{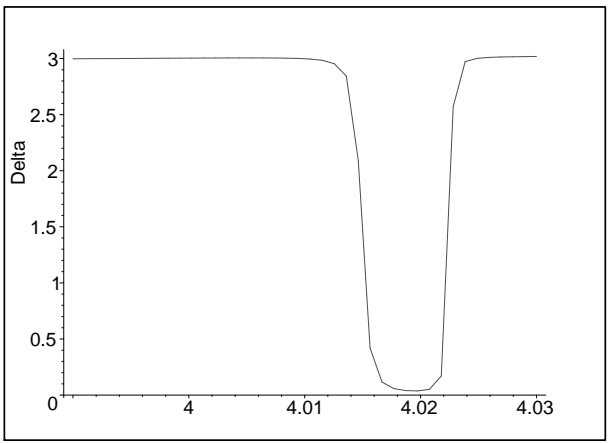}
\caption{\label{fig:fig5}$\Delta$ vs. $g^2Q^2_{ave}$.  This is for ordinal 
state 773.  The dip occurs where the state follows the trajectory of the 
transverse basis state.}
\end{figure}

This behavior is illustrated in Fig.~\ref{fig:fig5}.  $\Delta$ dips to 
a minimum less than 2\% of its asymptotic value.  Fig.~\ref{fig:fig4} 
confirmes that the dip occurs when $g^2Q^2_{ave}$ is in the interval where 
ordinal state 773 follows the transverse basis state trajectory. 

\begin{figure}
\includegraphics[angle=-90,width=8cm,keepaspectratio=true,
scale=0.5]{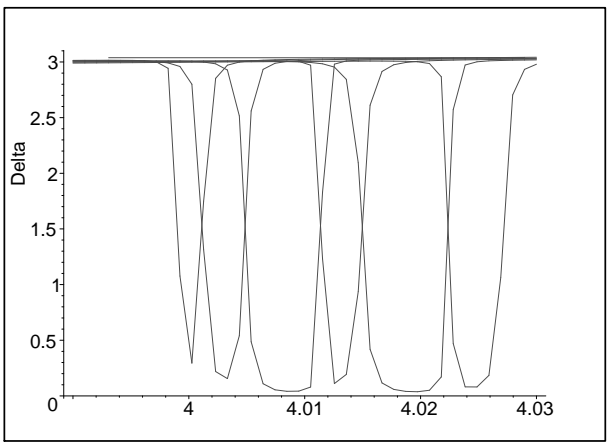}
\caption{\label{fig:fig6}$\Delta$ vs. $g^2Q^2_{ave}$.  Dips for ordinal 
states 772 (on right) through 777 (on left) are superposed.}
\end{figure}

As successive states follow the transverse basis state trajectory, the 
``locus of gauge invariance'' passes from one eigenstate to the next, always 
close to the transverse basis state.  This transition process is shown in 
Fig.~\ref{fig:fig6}, where the dips for states 772 through 777 are shown 
together.  Note that dips in $\Delta$ intersect where $\Delta$ is half its 
asymptotic value.  At an intersection, 
the normalized probability of finding the transverse basis state in the 
eigenstate, $|\beta_T|^2$, is about 1/2 for each of the ``dipping'' 
states.  (The sum of the probabilities over all the eigenstates of $H_B$ is 
one.)  At an intersection of dips, the ``locus'' passes from one eigenstate 
to the next.  

Thus, the ``locus of gauge invariance'' follows the transverse basis state.  
This conclusion is guaranteed by Eq. (\ref{eq:ampl}).  The extrapolation of 
this picture to $g^2Q^2_{ave}=0$ is provided by 
Table~\ref{tab:table1}.  This is as close as we can come to the 
verification that there is a second spin 0 gauge invariant state 
at energy 0.3843, thereby confirming double degeneracy of the spin 
zero gauge invariant states.  
\medskip
\section{Conclusion}

We now consider how the methods we have developed can be applied to 
pure SU(3) lattice gauge theory at general coupling.  In this case, all 
the terms in Eqs. (\ref{eq:electric1}) and (\ref{eq:ewe}) contribute to 
$H_A$.  The ``chromoelectric'' operators ${\cal J}^2$ each produce terms 
which are a product of up to four $A$'s and $A^\dagger$'s.  The 
``chromomagnetic'' operators ${\rm Tr}(UUU^\dagger U^\dagger)$ each produce 
terms which are a product of up to eight.  The operator $g^2Q^2$ produces 
terms which are a product of up to four such factors.  

Terms having more than two factors are interaction-type, and are not 
present in the weak coupling limit we have been studying.

The generator $G_B$ is now constructed from the terms in $H_A$ and $g^2Q^2$ 
that are no more than quadratic in $A$'s and $A^\dagger$'s, using Eq. 
(\ref{eq:generator}).  The basis is chosen from the low energy, one pair 
eigenstates of this generator.

The main complication beyond those we have studied comes from the fact 
that the expression for $H_B$, Eq. (\ref{eq:finalham}), now has 
interaction terms on the right side.  When these terms are added 
to $-\xi N$, the resulting interaction term is no longer a rank one 
operator.  Therefore, the construction of the eigenstates of $H_B$ is 
more complex than it is for the weak coupling case we have studied.

Interaction terms may pose a technical challenge, but they are required if 
our approach is to produce states like glueballs on the lattice.  To see 
this, note that Eq. (\ref{eq:momentum}) implies that a pair state is an 
excitation in which the amplitude for the excitation of two links has the 
spatial dependence $e^{2\pi i{\bf k\cdot\Delta s}/L}$, where $\bf\Delta s$ 
is the distance between the links.  The magnitude of this amplitude is 
independent of $|{\bf\Delta s}|$.  In a glueball, on the other hand, the 
amplitude falls off in magnitude when $|{\bf\Delta s}|$ is more than the 
diameter of the glueball.  Such behavior can be approximated when the 
eigenstates of $H_B$ are superpositions of many pair states, which in turn 
requires that there be interaction terms between pair states.

One concern is that there may not be enough basis pair states to superpose 
to make a glueball.  There are always many longitudinal polarization pair 
states in a basis, but that is not true for transverse pair states.  An 
extreme example is provided by the case we have used extensively in this 
paper, a lattice with $\eta=0.1$ and $g^2Q^2_{ave}\sim 4.0$.  For these 
parameters, there are 815 longitudinal polarization pair states and just one 
transverse pair momentum.  (There are two degenerate states, one for each 
transverse polarization.)  To get more transverse states in the basis, we 
increase $\eta$, which increases the energy interval between longitudinal 
pair states.  Table \ref{tab:table3} lists the number of longitudinal and 
transverse pair states for different values of $g^2Q^2_{ave}$, when 
$\eta=0.5$.  These results are for the weak coupling generator, but they 
suggest that it will be possible construct a basis with many transverse 
polarization states.
\begin{table}
\caption{\label{tab:table3}  Number of longitudinal (L) and transverse (T) 
seed momenta in bases having different $g^2Q^2_{ave}$.  $\eta=0.5$.}
\begin{ruledtabular}
\begin{tabular}{ccccc}
$g^2Q^2_{ave}$:&4.000&6.000&8.000&10.000\\
&&&&\\
\hline
L&108&113&113&112\\
T&4&9&18&30\\
\end{tabular}
\end{ruledtabular}
\end{table}

It should be remembered that adding states to a basis can only improve the 
accuracy of the approximate eigenvalues we compute.  Thus we are not obliged 
to slavishly follow the rules we have set forth for a basis. One can add 
further transverse pairs having energies greater than the highest energy 
longitudinal pairs to improve the ``localization '' of glueballs.  One of 
the attractive features of a variational basis is that ideas like this can 
be tried. 

These considerations set the stage for the exploration of the general 
coupling case, which is the goal of this line of research.  However, 
the exploration lies beyond the scope of this paper.

\end{document}